\documentclass[%
reprint,
superscriptaddress,
floatfix,
amsmath,
amssymb,
aps,
notitlepage
]{revtex4-2}
\usepackage[hidelinks=true]{hyperref}
\usepackage{xcolor}
\definecolor{greenish}{RGB}{35,162,18}
\definecolor{reddish}{RGB}{174,12,48}
\definecolor{blueish}{RGB}{14,40,178}
\hypersetup{
	colorlinks  = true, 
	urlcolor    = blueish, 
	linkcolor   = reddish , 
	citecolor   = greenish 
}

\usepackage{setspace}

\usepackage{tikz,pgfplots} 
\usetikzlibrary{external}
\pgfplotsset{
	compat=newest,
	invoke before crossref tikzpicture={\tikzexternaldisable},
	invoke after crossref tikzpicture={\tikzexternalenable},
}
\usetikzlibrary{
	external,
}
\pgfplotsset{plot coordinates/math parser=false} 
\usetikzlibrary{plotmarks}
\usepackage{caption}
\usepackage{siunitx}
\usepackage{subcaption}
\usepackage{graphicx}
\usepackage{dcolumn}
\usepackage{bm}
\usepackage{placeins}
\usepackage{gensymb}
\usepackage{float}

\def\phi{\varphi}

\begin{document}
\title{Learning to write with the fluid rope trick}
	
\author{Gaurav Chaudhary}
\affiliation{School of Engineering and Applied Sciences, Harvard University, Cambridge, MA 02138.}
\affiliation{Department of Mechanical Engineering, Massachusetts Institute of Technology, Cambridge, MA 02139}
    
\author{Stephanie Christ}
\affiliation{School of Engineering and Applied Sciences, Harvard University, Cambridge, MA 02138.}
    
\author{A John Hart}
\affiliation{Department of Mechanical Engineering, Massachusetts Institute of Technology, Cambridge, MA 02139}

\author{L Mahadevan}
\email{lmahadev@g.harvard.edu}
\affiliation{School of Engineering and Applied Sciences, Department of Physics, Department of Organismic and Evolutionary Biology, Harvard University, Cambridge, MA 02138.}
\date{}

\begin{abstract}
The range and speed of direct ink writing, the workhorse of 3d and 4d printing, is limited by the practice of liquid extrusion from a nozzle just above the surface to prevent instabilities to cause deviations from the required print path. But what if could harness the  ``fluid rope trick", whence a thin stream of viscous fluid falling from a height spontaneously folds or coils, to write specified patterns on a substrate? Using Deep Reinforcement Learning we control the motion of the extruding nozzle and thence the fluid patterns that are deposited on the surface. The learner (nozzle) repeatedly interacts with the environment (a viscous filament simulator), and improves its strategy using the results of this experience. We demonstrate the results in an experimental setting where the learned motion control instructions are used to drive a viscous jet to accomplish complex tasks such as cursive writing and Pollockian paintings.
\end{abstract}

\maketitle

\section{Introduction}
The rapid evolution of three-dimensional (3D) printing technology has enabled new manufacturing capabilities\cite{sydney2016biomimetic}, and typically involves layer-by-layer deposition of material through a computer-controlled nozzle. While motion control of the nozzle and development of materials with suitable rheological properties for printing have sped up the adoption of 3D printing, from a physical perspective, the ultimate limits on print quality are imposed by the fluid dynamics of the printed material. Such effects are seen practically in non-uniformly extruded/deposited material and instabilities such as folding and coiling of fluid jets \cite{yuk2018new}. A simple protocol to prevent these defects necessitates limiting the nozzle trajectory to exactly mimic the target print pattern from a very small height offset.

While this approach increases the accuracy of  layer-by-layer direct writing of 3D objects,  the result is that printing is typically a very slow process. Moreover, it is challenging to adapt these methods to complex topographies of the substrate on which the deposited material is laid down, or to precisely control sharp turns without distorting the extruded filament. Here, we explore the possibility that 3D printing could be sped up by operating with a nozzle that is elevated above the print surface, and harnessing the dynamic instability of a falling fluid jet \cite{barnes1958liquid, Mahadevan1998,Mahadevan2000,kim2010, ribe2017,chakrabarti2021} to enable rapid accurate printing without requiring the nozzle to exactly mimic the target pattern. This technique relies on finding printing solutions, i.e. nozzle paths in space-time that take advantage of folding and coiling instabilities rather than avoiding them. Thus, given a final target pattern or shape in a plane, we ask how to optimize the nozzle trajectory so that a continuous stream of material can be printed without having to precisely translate the nozzle along the print path, or bringing the nozzle close to the substrate. Proof that this can be done, and very well, is seen in the striking  art of Jackson Pollock, who created paintings by dripping and pouring paint on a canvas from a height while moving his hand  \cite{Herczynski2011}. Can a machine be trained to learn this technique?



Our approach taps into the recent success of a class of machine learning algorithms known as reinforcement learning methods \cite{Sutton2017} that are able to harness the  expressive power of neural nets (NN) to explore and exploit a large state and action space to find solutions to difficult tasks. Our starting point is a physical simulator that characterizes the necessary physics of thin threads of viscous fluid extruded from a height and captures the interactions between agent (nozzle head) and environment (print surface). Then, by building a framework of learning through repeated interactions with the environment, we show that it is possible to find a nozzle trajectory that can print a target pattern.


\section{The dynamics of viscous filaments}

\begin{figure}[htp!]
    \centering
	\includegraphics[width = 0.4\textwidth]{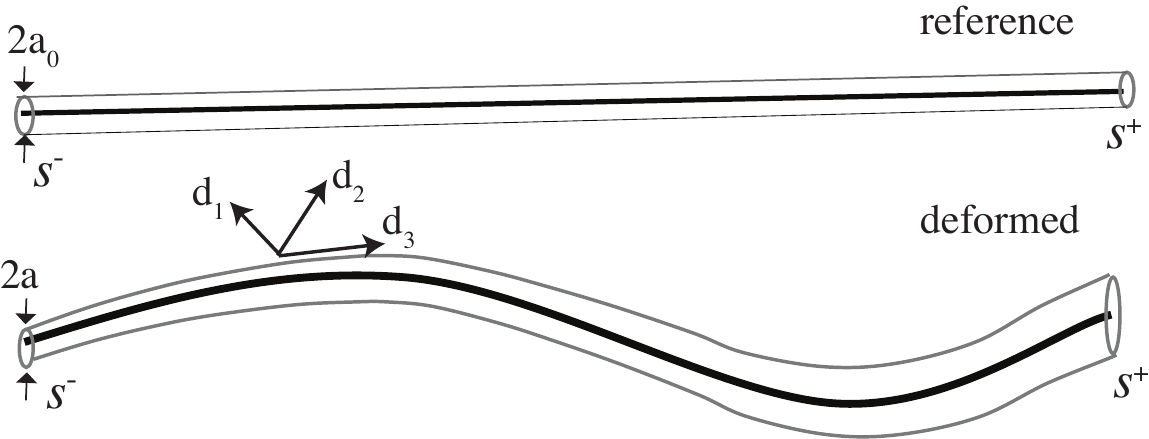}
	 \caption{Representation of the fluid thread and the associated material frame.}
	\label{fig:figure_centerline}
\end{figure}

\begin{figure*}[htp!]
    \centering
	\includegraphics[width = 0.7\textwidth]{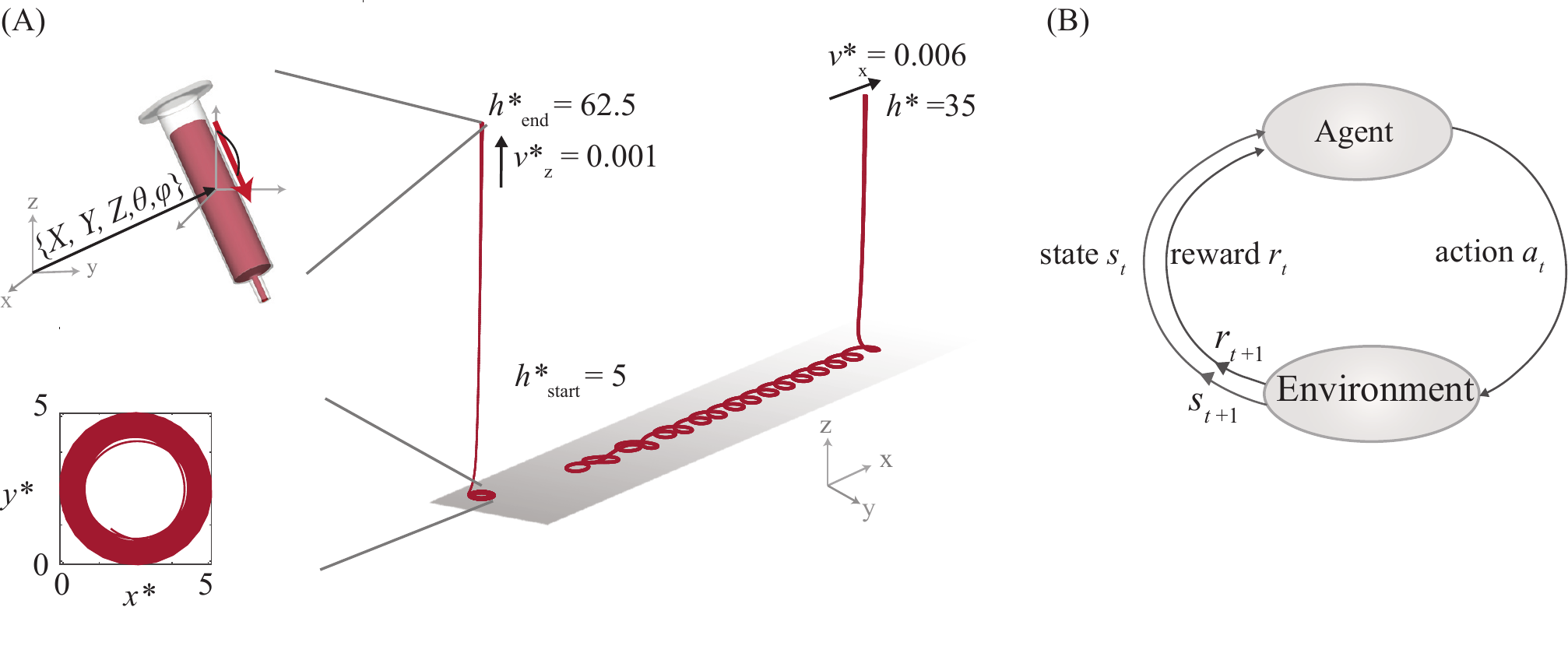}
	 \caption{\textbf{Printing with the liquid-rope instability}. (A) The state of a printing nozzle is represented by its translational coordinates $(X, Y, Z)$ w.r.t. a stationary frame. The azimuthal angle $\phi$ and polar angle $\theta$ further specify the ink extrusion direction (solid red arrow). Simulation of viscous thread ($\eta = 5000$~cP) coiling as the print head is translated vertically at a constant speed. The top view shows coils of increasing radii as the nozzle is raised from $z=0.5$ to $z=12.5$~cm. Translating the nozzle at a constant speed of $0.5$~cm/s creates periodic coils on the surface as shown for a fixed set of parameters. (B) In the controlled setting, the agent (nozzle) interacts with the environment to improve its action-selection policy $\pi$. It receives a state $s_t$ and numerical reward $r_t$ from the environment, based on which it updates its policy and returns the next action $a_t$. By taking action $a_t$, the environment transitions to the next state $s_{t+1}$. It measures its desirability through the reward $r_{t+1}$ and sends the information, along with the new state, back to the agent.}
	\label{fig:figure_1}
\end{figure*}

The simulator used to model the liquid-rope instability of a viscous fluid follows earlier work on characterizing the coiling of liquid jets\cite{Audoly2013}, which we summarize briefly.  A thin stream of viscous thread is represented using a discrete Lagrangian description that accounts for the narrow aspect ratio (length/radius) of the viscous thread; then the centerline representation of the coordinates provides a reduced order description for the fluid threads. The centerline $x(S,t) \in {R}^3$ and the material frame $[d_1(S,t), \,d_2(S,t), \,d_3(S,t)]$ is used to capture the time evolution of viscous thread as shown in Fig.~\ref{fig:figure_centerline}. The material frame is adapted to centerline by requiring $d_3$ to align with the tangent to the centerline, and $d_1, d_2$ span the plane normal to centerline's tangent. 

The dynamical equations of the thread per unit length of the thread are given by the balance of linear and angular momentum for the position $x(S,t)$ and the spin component of the angular velocity at every cross-section $v(S,t) $\cite{Audoly2013}:
\begin{align}
    \rho \, A_0 \,\ddot{x}(S,t) &= F_v(S, t) + f(S, t) \\
    l \, J \, \dot{v}(S,t) &= M_v(S, t) + m(S, t)
\end{align}

Here, $F_v$ and $M_v$ are internal resultant viscous forces and moments in the thread arising from the rates of stretching, bending and twisting of the thread, while $f$ and $m$ are the external body force and body moment density, respectively. The expressions for the resultant force and moments in the fluid thread are expressed using the kinematics of the centerline representation in terms of the local area $A_0$ and moments of inertia $J$ at every cross-section $S$, as well as the  viscosity of the fluid (assuming it be modeled as a simple Newtonian liquid);  we refer the reader to \cite{Audoly2013}, \cite{ribe2017} for the detailed expressions of the individual terms in the above equation and their respective discretizations, which form the basis for this work \footnotetext{\ddag~Codebase: https://gitlab.com/stchrist/ViscousThreads}.

A critical dynamical variable underlying the generation of complex patterns \cite{Morris2008} is the ability to move the nozzle; we assume that it can be moved freely in a plane using horizontal movements and also allow it to rotate relative to two axes, as shown in Fig.~\ref{fig:figure_1}~A, with the angles $\phi$ and $\theta$ describing the rotational degrees of freedom (DoF).  Assuming that the flow rate ($Q$) and nozzle extrusion diameter $a$ are constant, controlling the writing/painting pattern on the substrate requires that the agent (nozzle) have time-varying translational and rotational velocities, likely to be bounded within a range of practical relevance. Vertical motion of the nozzle changes the nature of the pattern deposited; at low heights, the jet is rectilinear, then starts to coil as the height is increased past a first threshold, and then the jet coils rapidly as inertial effects start to dominate at even larger falling heights. In Fig.~\ref{fig:figure_1}~A (bottom), we show a coiling pattern  deposited on the surface ($z=0$); the coil radius increases as the height of the nozzle is increased. When the agent is translated linearly along the $y-$ axis with a constant velocity, following a short transient, we see that a coiling pattern is deposited on the substrate $z=0$ similar to what has been observed experimentally \cite{Morris2008}. 
These examples suggest that to write a desired pattern with such an agent, we have to actively control the various DoFs of the nozzle. But how can we learn the form of the agent's actions?

\section{A reinforcement learning framework}
Reinforcement learning (RL) \cite{Sutton2017} provides one possible answer by allowing the action of an agent to repeatedly interact with the environment and converge towards an optimal policy without having any prior knowledge of the underlying model.  An RL problem is defined in terms of states, actions and rewards. The state $s \in S$ is a quantitative description of the environment at the current time, with $S$ being the set of all possible states. In every state, a set of actions $A(s)$ are available to the agent. By taking an action $a \in A(s)$, the agent transitions from one state to another and receives a numerical scalar reward signal $r$ from the environment. The reward is a measure of how desirable it is to take action $a$ in state $s$. Overall, the goal of the learner (or agent) is to find an action policy $\pi(a|s)$ which maximizes the cumulative reward over the entire learning episode by exploring different ways of interacting with the environment; this schematic is summarized in Fig.~\ref{fig:figure_1}~B.

For the printing problem, the current state is defined by the following continuous parameters: the arc length $S$ of the pattern that has been printed so far, the current position of the nozzle $(x, y, z, \phi, \theta)$ with $z \in [h_{\text{min}}, h_{\text{max}}]$, and the last action, corresponding to the motion of the nozzle $(v_x, v_y, v_z, v_\phi, v_\theta)$. If the height of the nozzle $z$ is outside of its prescribed range, the state is considered invalid. Just as the states, the actions (DoFs) that describe the motion of the nozzle  are continuous. The agent has up to five degrees of freedom: translation velocities ($v_x, v_y, v_z$) and rotational velocity ($v_\phi, v_\theta$). Furthermore, we assume that the range of all actions is bounded within the interval $\left[ v_{\text{min}}, v_{\text{max}} \right]$. For the position and motion of the nozzle, only the DoF that are part of the action are considered, as the other parameters stay constant. Thus, the dimension of the state is between 3 (one DoF) and 11 (five DoF). Theoretically, it would be sufficient to use the arc length as the only only state variable, however, we find that the rewards converge very slowly in that case.

More specifically, we use an off-policy actor-critic named V-RACER \cite{novati2019remember} as our RL framework, although many other methods are likely to also work. A summary of this follows: the algorithm trains a neural network to approximate a continuous path $\pi^{w}(a|s)$ for the nozzle (continuous policy approximation). The policy network is randomly initialized and then iteratively updated through repeated attempts to reach the target following the policy gradient theorem. We employ "Remember and Forget Experience Replay" to reuse past experiences over multiple iterations to update the policy in a stable and data-efficient manner, with hidden network layers with 128 LSTM (long short-term memory) units each. This representational capacity of the network was found to be sufficient for the cases considered in this work. The actuation period (temporal resolution) of our simulations was generally fixed to 0.01 s; varying the number of units and the number of layers did not change the final optimized solutions. 

Since the goal of the learning to write is equivalent to depositing a 2D pattern of the same shape as a target input pattern, the reward/penalty is chosen so as to reflect the mismatch between the target and the printed pattern. The target pattern is given as an ordered list of 2D coordinates. We implemented the reward by comparing the position of the printed vertices with vertices of the given target shape. Denoting the total arc length of the pattern that has been created so far by $S_i$ -- where $i$ is the current learning step, the error is computed by integrating it from the last step to the current step:
\begin{equation}
	r = - \int_{S_{i-1}}^{S_i} \lvert \tilde{\mathbf{x}}_{target}(S) - \tilde{\mathbf{x}}(S) \rvert \, dS
\end{equation}
Here we have chosen the error at a certain arc length in terms of the absolute difference between the target position and the actual position, linearly interpolated from their two respective neighbors. 
In addition to the reward based on the pattern, we give a strong negative reward for invalid states, i.e. for invalid heights $z$. Using this reward function, the maximum -- and optimal -- cumulative reward is 0, as this would indicate that the created pattern does not deviate from the target pattern at any point.



To create regular initial conditions for learning, the simulation is first stabilized by using a constant velocity  $(v_x, v_y, v_z, v_\phi, v_\theta) = (v_0, 0, 0, 0, 0)$ for a fixed number of time steps until the pattern follows a simple straight line. From there, the agent is allowed to start learning the appropriate set of actions to replicate the target pattern. In practice, in addition to the simulation parameters (nozzle flow/radius and fluid properties) and RL parameters (learning rate, discount factor, size of NN, activation function), we also specify the maximum number of steps per episode step $n_s$, i.e. the number of actions that the agent can take before the episode is over.

\section{Using reinforcement to learn writing}

\begin{figure*}
    \centering
    \includegraphics[width=0.7\textwidth]{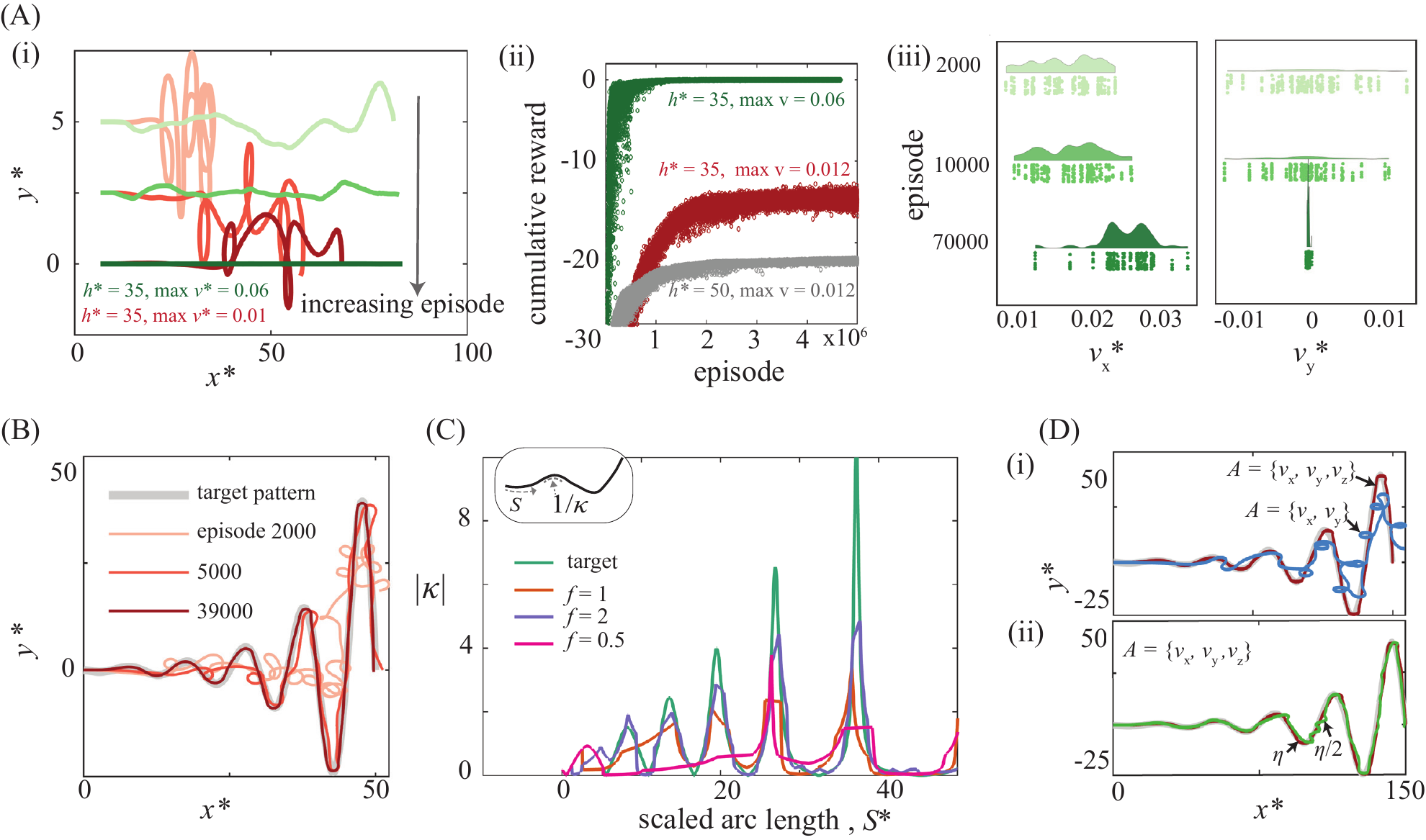}
    \caption{\textbf{Using reinforcement learning to control the liquid coiling instability}. (A-i) The agent takes actions, $a_t =\{v_x, v_y\}$ while extruding liquid from a fixed height of 7~cm and learns to draw a straight line over multiple iterations, seeking to maximize the total reward. The red and green curves correspond of the maximum limit of 1~cm/s and 5~cm/s on the velocity components. (A-ii) Maximum reward depends on the limits on translation velocity, and the height from which the liquid is extruded. A better control can be learned when extruding from a smaller height for the same limits of velocity components. (A-iii) A density plot of the actions taken at various episodes. To create a straight line, the agent learns to suppress coiling by moving solely in the x-direction. (B) A more complex task of drawing a varying amplitude wave using a larger set of actions, $a_t =\{v_x, v_y, v_z\}$. Unlike the straight line in (A), the agent harnesses the liquid coils to draw the curved pattern. (C) The performance of the agent also depends on the relative scale of liquid coils and target curvature. A comparison of the local curvature ($|\kappa|$) of the drawn pattern at different scales show the challenge in learning when the scale of target curvature is significantly smaller than the nature curvature of liquid coils. (D-i) The constrained set of actions limits the agent's performance in drawing larger curvature. (D-ii) The fluid material properties, such as the kinematic viscosity $\eta$, also limit the accuracy of the learned pattern.}
    \label{fig:figure_2_line_sinewave}
\end{figure*}


To validate that the RL agent takes optimal actions, we first ask if we can reproduce the simplified scenario of printing a straight line from a nozzle. As shown earlier in Fig.~\ref{fig:figure_1}~A, for an arbitrary unidirectional velocity, an uncontrolled agent will produce a pattern consisting of overlaid coils. If the nozzle is allowed to vary its planar velocity ($v_x$ and $v_y$), we see that over time associated with an increase in the number of episodes, the agent converges to create a straight line as shown in Fig.~\ref{fig:figure_2_line_sinewave}~A~i. Consistent with this, the cumulative reward plateaus with increasing episode number (Fig.~\ref{fig:figure_2_line_sinewave}~A~ii); the value of the plateau depends on the height of fluid extrusion and bounds on the nozzle actions. To draw a straight line, the agent speed must counter the effective velocity at which the coils are naturally laid on the surface, i.e. the optimal solution must satisfy the relations $v_x \ge \sim \Omega R$, and $v_y \sim 0$. The action density plots in Fig.~\ref{fig:figure_2_line_sinewave}~A~iii show that the agent does indeed converge to this state. 

To truly learn to write, the agent must learn to negotiate curves with complex curvature profiles. Therefore,  we need a target pattern with a wide range of curvatures and rate of variation of curvature; an exponentially decaying sinusoidal wave $ y = 10 \, e^{-1000\, x} \,\sin {2000\pi x}$ serves both purposes. The agent is now allowed an additional action $v_z \ne 0$ in addition to the ability to vary $v_x, v)y$, and in Fig.~\ref{fig:figure_2_line_sinewave}~B we show the learned patterns as a function of episode number, and see that after about 5\,000 episodes, the correct shape is learned, and the visible accuracy of the task does not improve much beyond that. A closer look at the evolution of the learned pattern shows that the agent first learns to match the pattern on larger length scales, followed by further improvements to closely match the features with sharp changes in curvature, i.e. on smaller length scales. This is consistent with the cumulative reward first showing a sharp increase followed by a very weak increase corresponding to the plateau-like regime in Fig.~\ref{fig:figure_2_line_sinewave}~A~ii) associated with later episodes. 

An optimal strategy for the agent would aim to utilize or avoid the coiling instability depending on the nature of the curvature profiles in the target pattern. Thus, for curvatures comparable to the coiling radius, it makes sense to use the natural coiling instability, and otherwise to avoid it by moving quickly in the plane. However, when the (position and velocity) actions of the nozzle are bounded, writing patterns with curvatures that are significantly larger than the coil radius can be a challenge.  To challenge the agent with such patterns, we change the absolute scale of the target pattern, thus changing the curvature profiles overall, and ask how well the RL agent, with the same bounds on action, learns to draw patterns at different scales. In Fig.~\ref{fig:figure_2_line_sinewave}~C we compare the absolute curvature of the  target pattern with that of the learned patterns as a function of scale, using the same exponential form used in Fig.~\ref{fig:figure_2_line_sinewave}~B. For each case, the overall scale of the target pattern is halved or doubled, keeping all other parameters fixed. We see that the magnitude of curvature of the learned patterns is generally different from the target curves at the locations of extreme curvature; the case with smallest scales shows significant deviations from the target, implying that decreasing the scale of the target pattern (hence increasing the curvature), results in a reduction of the agent's ability to print the target, consistent with intuition.

The substrate patterns associated with the jet coiling instability are affected by both the limits on the dynamics of the nozzle as well as the properties of the fluid. To explore their respective roles, in Fig.~\ref{fig:figure_2_line_sinewave}~D~ii we show the effect of limiting the agent's actions to {$v_x, v_y \ne 0, v_z =0$}, and compare it to the case when $v_z \ne 0$ (Fig.~\ref{fig:figure_2_line_sinewave}~B); we see that extruding fluid from a constant height introduces an unwanted coiling response in the learned pattern, especially near the regions of large curvature in the target pattern, demonstrating the influence of limiting the action space in the learning problem.
Fig.~\ref{fig:figure_2_line_sinewave}~D~ii shows the effect of changing the magnitude of the viscous forces by simply changing the kinematic viscosity ($\eta$) of the fluid; reducing the viscosity leads to an increase in the coiling frequency in a predictable way and we see that after a similar number of learning episodes, the learned pattern is not quite as accurate. These tests of deploying RL point to a simple but important lesson; choosing the right range of actions is critical for good performance in any task, and becomes particularly clear in the context of interacting with and learning using physical systems.  

\begin{figure*}
    \centering
    \includegraphics[width=0.7\textwidth]{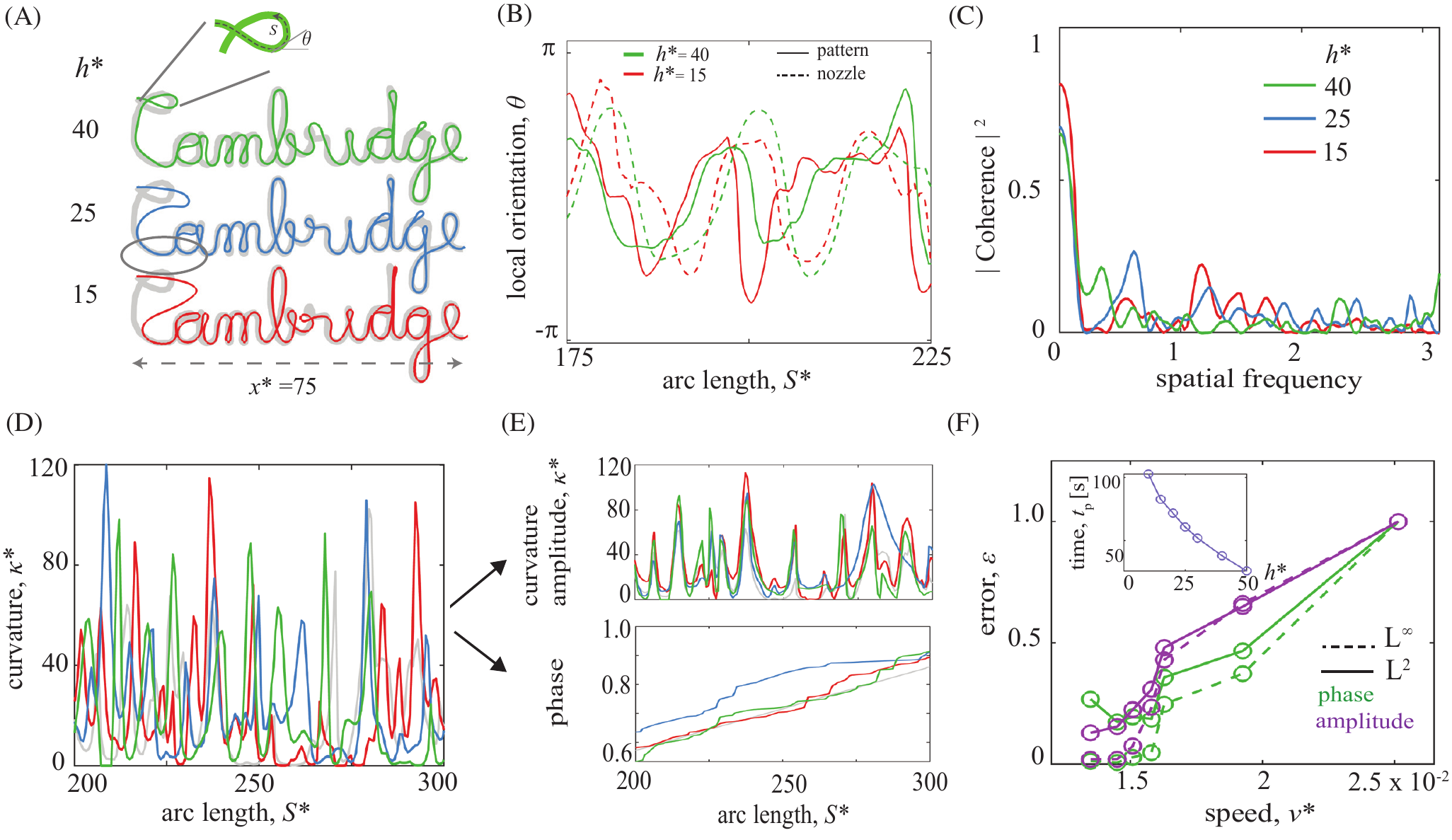}
    \caption{\textbf{Using RL to reproduce cursive writing}. The agent, with actions $a_t =\{v_x, v_y\}$, can harness liquid coiling to write cursive text. (A) The predicted deposited patterns from fixed heights as shown. In each case the z-component of velocity is set to zero. (B) A comparison of local direction (local orientation w.r.t the horizontal $\theta$) of translation of the nozzle and the printed pattern show that non-trivial actions are performed to control the target pattern, and the agent does not just follow the trajectory of target pattern. (C) The magnitude squared coherence between the printed patterns and the target pattern is shown. A higher coherence is observed at lower spatial frequencies.
    (D) The curvature of the cursive text plotted as a function of arc length shows misaligned curvature peaks for patterns deposited from different heights. The error in $\kappa$ vs $S$ curves can be decomposed into vertical and a horizontal components (in E) using elastic functional data analysis. (F) Changing the height of extrusion allows for directly controlling the total printing time (inset). A speed-error trade-off results as a consequence. Within a given range of actions, a lower error, defined as $\varepsilon = (\int (1 - \kappa(S) / \kappa_0(S))^p dS)^{1/p} $, is achieved when nozzle is held closer to the surface as a result of low-coiling frequency and smaller liquid coil radii. Here we define non-dimensional printing speed $v^* = (S_{total}/t_p)/\sqrt{g h}$}
    \label{fig:figure_3_cursive}
\end{figure*}

Following the calibration of the ability of the RL agent, a moving nozzle, to learn to write out patterns suggests that it is possible to print any continuous planar curve, given a sufficiently large and bounded action space. To further explore the ability of RL in demonstrably difficult tasks, we now demonstrate that RL can be used to guide a nozzle to mimic cursive writing or paint like Pollock. We first prescribe the target pattern: a continuous curve associated with the word ``Cambridge" along with an action space $(v_x, v_y)$, but no vertical velocity, i.e. $v_z=0$. In Fig.~\ref{fig:figure_3_cursive}~A, we show how the agent learns to write from different fixed heights in the range $z \in [2,10]$~cm, after the same number of training episodes. While the overall patterns look similar, a closer look shows noticeable differences, especially in the regions of the curve with large curvature e.g. letters ``r", ``d" and ``g".  Since the coil radius and frequency depend on the nozzle height, the actions taken depend on this parameter. In regions of the pattern that are almost straight (or weakly curved), the agent avoids the coiling instability by moving rapidly; similarly, it uses the coiling instability to achieve higher curvature in regions where the target pattern demands it. 

To understand the discrepancies between the target and learned patterns, in Fig.~\ref{fig:figure_3_cursive}~B we show the local direction of motion of the agent and local direction of the pattern that gets laid on the surface. It is clear that there are two types of errors: at some places, the local orientation of curves appears phase-shifted, while in other regions the amplitudes of the curves are poorly correlated. 

To quantify the accuracy of learning the print path, we take two approaches that use local measures to inspire global metrics, noting that since we work with planar curves, upto rigid motions (that we do not worry about), the target and learned patterns are completely characterized by their scalar curvature as a function of the arc-length. For one measure of the error on the scale of the whole pattern, we define a coherence metric in the spatial Fourier domain as follows : 
\begin{equation}
C_{\kappa_0\kappa}(f) = \frac{|P_{\kappa_0\kappa}(f)|^2 }{ (P_{\kappa_0\kappa_0}(f)\,P_{\kappa\kappa}(f))}
\end{equation}
where $P_{\kappa\kappa}(f)$ and $P_{\kappa_0\kappa_0}(f)$ are the power spectral densities of the curvatures $\kappa_0(S)$ (target) and $\kappa(S)$ (learned), respectively, and $P_{\kappa_0\kappa}(f)$ is the cross power spectral density between $\kappa_0(S)$ and $\kappa(S)$. \cite{kay1988modern}, we note that $C_{\kappa_0\kappa}(f) \in [0, 1] $. Plotting the coherence in Fig.~\ref{fig:figure_3_cursive}~C, we note that $\kappa_0(S)$ and $\kappa(S)$ are strongly associated at very low spatial frequencies (or relatively larger scales) while at higher spatial frequencies the association is relatively weak. This is consistent with the observations from other learning experiments as well: the agent is can capture the large scale features of the curves either by moving in straighter or slightly curved path or printing coils, but when the target curvature scales are larger than the natural coiling scales, a mismatch between the target and learned curvature may occur. 

To compare the difference between the target and learned curvature along the trajectory raises a familiar problem of registration - how does one align points along the learned path with points along the target pattern, given that the speed of the nozzle is not necessarily a constant? To solve this problem, we use methods from functional and shape data analysis \cite{srivastava2011registration} whereby we simultaneously solve the problem of registration and determination of error by insisting on reparametrization invariance of the error metric. A natural solution that presents itself is to align the curvature data for various cases with the curvature of the target curve using the Fisher-Rao metric and the square-root velocity function (SRVF) representation of curvature data \cite{srivastava2011registration}, thus separating out errors in the phase and the amplitude of the curvature. Fig.~\ref{fig:figure_3_cursive}~E shows the aligned curvature of different cases shows in Fig.~\ref{fig:figure_3_cursive}~D and their respective phase shift. Defining the error as the $l-p$ norm of the curvature mismatch, 
\begin{equation}
\varepsilon = \left[\int \left(1 - \frac{\kappa(S)} {\kappa_0(S)}\right )^p dS \right]^{1/p} 
\end{equation}

\begin{figure}
    \centering
    \includegraphics[width=0.475\textwidth]{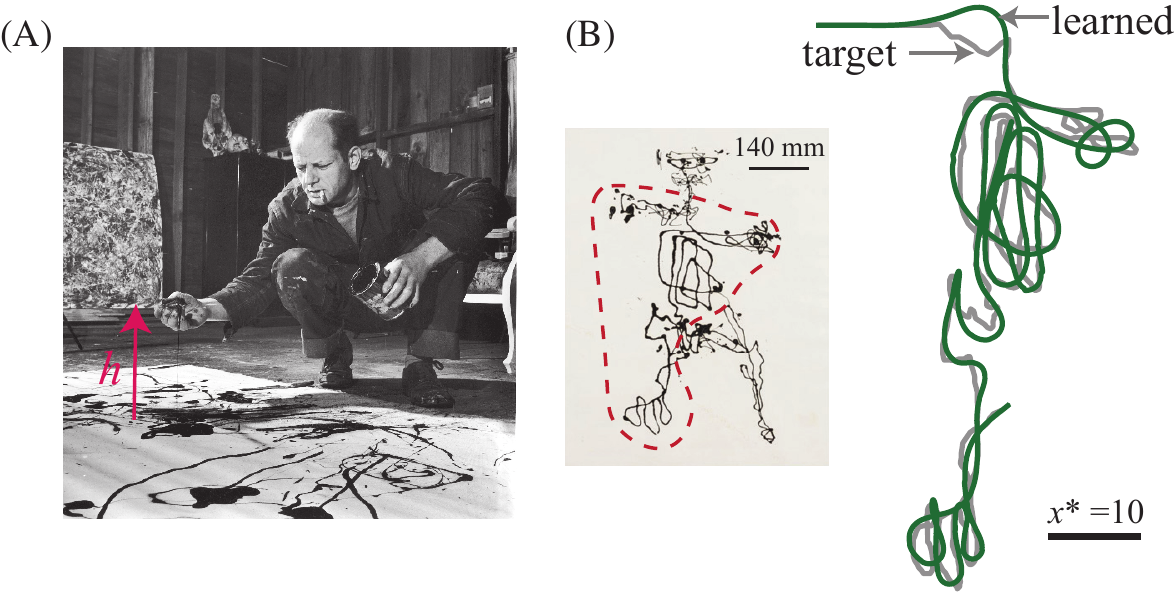}
    \caption{\textbf{Using RL to mimic Pollock's paintings}. Liquid coiling was extensively utilized by famous American painter Jackson Pollock in his drip paintings (as shown in (A) \cite{pollock_1970}). (B) With the available set of action $(v_x, v_y, v_z)$, the agent can learn to draw parts of Jackson Pollock's, \emph{Figure, 1948}, indicating that Pollock's drip painting may owe its complexity to the liquid coiling instability.}
    \label{fig:figure_4_pollock}
\end{figure}

Based on the error defined above, we quantify agent's capability to print from various heights above the surface. 
Printing with different nozzle heights is associated with a trade-off. Small nozzle heights lead to very accurate learned paths but ones that are very slow, since they do not exploit the instability, while larger nozzle heights lead to faster printing albeit with poorer accuracy. In the inset of Fig.~\ref{fig:figure_3_cursive}~F we quantify show that a higher extrusion height results in a higher coiling frequency and hence faster overall printing. In Fig.~\ref{fig:figure_3_cursive}~F we show the trade-off by comparing the overall curves printed from different heights; faster printing is also less accurate.  We note that there is an additional effect, since the filament diameter at the substrate changes with the height of the nozzle, but here we ignore this and only consider matching the target pattern with the centerline of the fluid jet.

\begin{figure*}
    \centering
    \includegraphics[width=0.9\textwidth]{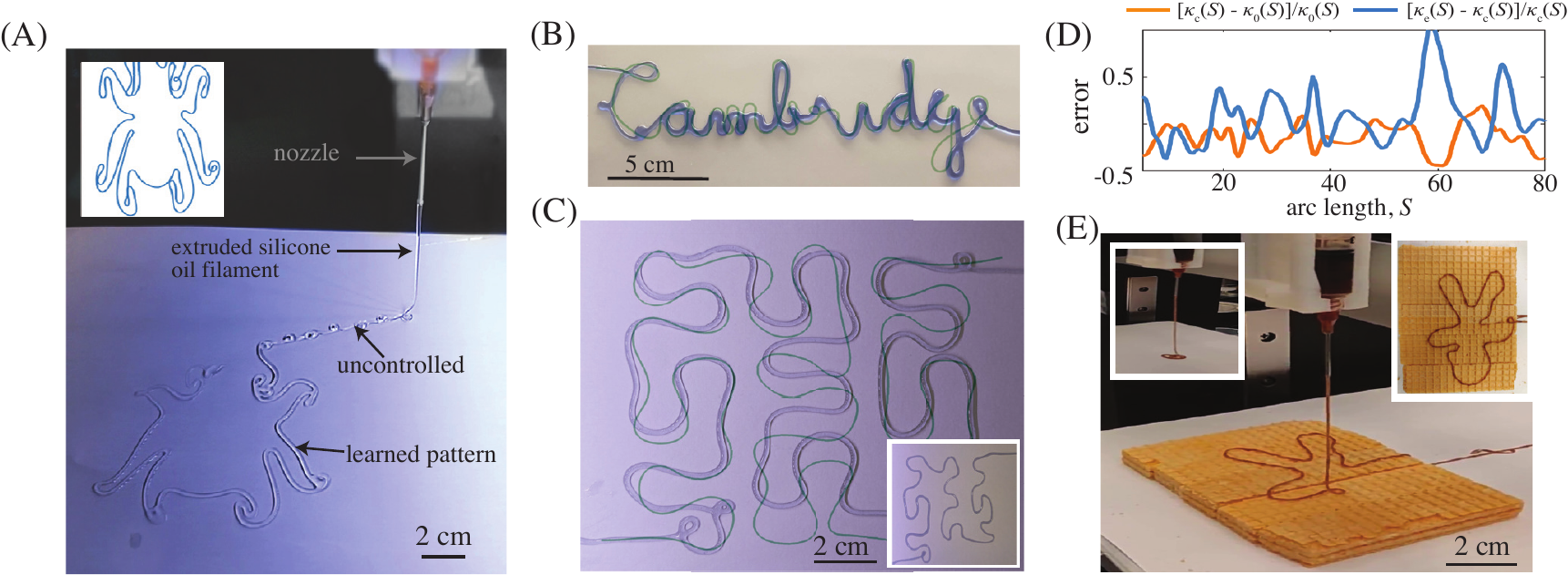}
    \caption{\textbf{Testing the predicted RL strategies using physical experiments to write}. (A) An experimental print showing a pattern with a first region printed with RL-derived and a second region without control where the nozzle is moved at 0.5~cm/s. (inset) The pattern printed in simulation. (B) RL controlled cursive handwriting using silicone oil, overlaid with the numerically simulated solution. (C) An experimentally printed self-similar structure (solid line), and the numerically learned pattern (dashed line). The mismatch between the experiment and numerical result grows as the arc length grows starting from the bottom left corner. (Inset) The experimental pattern deposited with a noisy set of actions shows the sensitivity of the final pattern to the RL actions. (D) A comparison of error in the curvature along the entire arc length of the pattern in (C) is shown. Here, $\kappa_c(S)$ and $\kappa_e(S)$ are the curvatures of computed and experimentally printed pattern, respectively, from the learned set of actions.(E) In a practical setting, the RL agent can be deployed to automate processes such as printing chocolate syrup on edible surfaces. (insets) a natural coiling of chocolate syrup extruded from a stationary nozzle held at a height of 6.6~cm above the surface and the top view of the final print.}
    \label{fig:figure_5_expt}
\end{figure*}

Finally we ask if the agent can learn to not just write, but sketch, paint or draw. As an exemplar, we chose part of a painting by Jackson Pollock (\emph{Figure}, 1948 \cite{pollock_1970})), famous for his "painting at a distance" style, wherein he allowed paint to drip and drizzle from a brush or a rod held far above the canvas, thus exploiting both fluid coiling and jetting instabilities \cite{Herczynski2011} as shown in Fig.~\ref{fig:figure_4_pollock}~A. Could an agent learn the complex movements of the artist given the final result?  Restricting ourselves to a limited part of the painting which we could easily track as a continuous curve as shown in Fig.~\ref{fig:figure_4_pollock}~B inset, we show that by allowing the action space for the RL agent to be non-zero {$v_x, v_y, v_z$} does in fact reproduce a fragment of the Pollockian painting as shown in Fig.~\ref{fig:figure_4_pollock}~B.

\section{Physical experiments testing how to learn writing}

The ultimate test of the RL protocols described above is to use physical experiments based on the learned strategies. To do so, we dispense a viscous fluid, silicone oil ($\eta=500$~cP), through a nozzle under a controlled flow rate, and deposit it along a digitally defined path on a smooth sheet of paper glued on a flat, stationary platform. We use the RL control that was learned in the numerical setting as the digital path along which the printer head translates. Since a fixed time step, $\Delta t = 0.01$~s was used in the simulations, we find this digital path by integrating the action set (velocities) at every step. The diameter of the nozzle and flow rate are same as the numerical experiments. In all the experiments the oil is first extruded until the coiling becomes steady, to ensure similar initial conditions as that of the numerical simulations.

We start by showing the difference between the RL protocol and uncontrolled flow, in Fig.~\ref{fig:figure_5_expt}~A (see S.I. video 1) for an experimental print shown in the inset. Using the learned RL control, the agent is able to smoothly print the pattern whereas when the agent moves at a constant velocity, fluid coiling readily occurs, thus producing a mimic of the given target pattern. To test the ability to write cursively, in Figure~\ref{fig:figure_5_expt}~B (see S.I. video 2) we show an example parallel to the simulation results in Fig.~\ref{fig:figure_3_cursive}~A using the action set {$v_x, v_y, v_z$}, with RL controls based on allowable states for nozzle heights $z \in [4-10]$ ~cm. An overlay of the numerically learned pattern on the experimentally printed pattern shows that the regions of highest errors are generally the regions of maximum curvature, most likely due to velocity mismatch between the RL actions and the experimental implementation. In particular when the nozzle makes sharp changes in direction over a short time duration (typically nearly high curvature regions), the executed (experimental) velocity actions are expected to be different from the desired actions. As a consequence, the overall mismatch (error) between the numerical and the experimental pattern builds up as a function of arc length of the pattern. We emphasize this using a self-similar target pattern of a Peano curve in Fig.~\ref{fig:figure_5_expt}~C (see S.I. video 3) using actions {$v_x, v_y$}. We observe that the experimental print and the simulated print show close agreement in the initial part of arc length, but as the printing progresses, the apparent experimental error accumulate over the arc length. A comparison between the curvatures of computed pattern and the target pattern (computational error), and experimentally printed pattern and computed pattern is shown in Fig.~\ref{fig:figure_5_expt}~D. In both the cases the error is computed between the aligned curvature vectors using the methodology described in the previous section. It is evident that along the entire arc length, the error in the curvature of the computed pattern deviates consistently from the target curvature. The error between the computed and the experimentally printed pattern has even larger magnitude due the additional experimental errors. In the inset of Fig.~\ref{fig:figure_5_expt}~C we show the sensitivity of the resulting pattern on the error in optimal set of actions. Here, we add a random noise to the actions ($< 5\%$ of the action value), and use the noisy actions to the print the pattern. The resulting pattern deviates significantly, highlighting the non-trivial nature of the action-reward landscape due to the unsteady and nonlinear effects in the physical problem.

As a test of the RL control approach developed for Newtonian fluids extruded on smooth surfaces, we ask how well the strategy does when printing  a thick chocolate syrup on a textured wafer, inspired by a tasty application of our approach. Chocolate syrup is a non-Newtonian fluid with a strain-rate dependent viscosity, bu here we assume that it has a constant shear viscosity ($\sim 100$~cP based on ball drop experiments). We find that this approximation results in a reasonable agreement between the printed shape and the target pattern as shown in Figure~\ref{fig:figure_5_expt}~D (see S.I. video 4). We note that printing from a height naturally handles rough surfaces, unlike the traditional direct-ink write where an irregular surface will result in irregularities in the deposited material.

\section{Conclusions}

A natural next step that follows understanding and predicting fluid instabilities is to control them. Inspired by 3-d and 4-d printing technologies that rely on the movement of a nozzle that dispenses complex fluids onto a substrate from just above, we ask if we can harness the folding and coiling instabilities that arise as soon as the jet falls from a sufficient height above a surface.  We answer this in the affirmative by combining a physics-based simulation engine  and a variant of reinforcement learning to control the fluid coiling instability and learn to "print at a distance."  By varying the action space, material properties and geometric scales that govern the dynamcis of viscous coiling, we quantified the performance of the RL agent for a variety of problems, and showed that it is possible to learn to write cursively and mimic Pollockian paintings. Deploying the learned policy in physical experiments, demonstrated that our approach can create complex physical patterns leveraging a natural fluid instability. We envision such an approach can be further extended to more challenging scenarios such as printing on non-planar surfaces and using robotic manipulators with greater dexterity and improved motion control.

\section*{Acknowledgements}	
The computations in this paper were run on the FASRC Cannon cluster supported by the FAS Division of Science Research Computing Group at Harvard University. We thank Crystal Owens and the Hatsopoulos Microfluids Laboratory at MIT for help and support with the experiments, and the  NSF Harvard MRSEC DMR-2011754, the Simons Foundation and the Henri Seydoux fund for partial financial support.





%

\end{document}